\documentclass[12pt]{article}
\usepackage{epsf}

\title{
 Effects of Matter Density Fluctuation in Long Baseline Neutrino
 Oscillation Experiments
}

\author{
 Masafumi Koike$^{a}$%
  \thanks{e-mail address: \tt koike@icrr.u-tokyo.ac.jp}
 and Joe Sato$^{b}$%
  \thanks{e-mail address: \tt joe@hep-th.phys.s.u-tokyo.ac.jp}\\
  {\small
   {\it $^a$ Institute for Cosmic Ray Research, University of Tokyo,}
  }\\
  {\small {\it Midori-cho, Tanashi, Tokyo 188-8502, Japan } }
  \\
  {\small {\it $^b$ Department of Physics, University of Tokyo,} }\\
  {\small {\it Bunkyo-ku, Hongo, Tokyo 133-0033, Japan } }
}

\date{}

\begin{document}
\maketitle
\abstract 
The effects of matter density fluctuation in long baseline neutrino
oscillation experiments are studied.  Effects of short wavelength
fluctuations are in general irrelevant.  Effects of long wavelength
fluctuations must be checked on a case-by-case basis.  As an example
we checked the fluctuation effects and showed its irrelevance in a
case of K2K experiments.
\vspace{0.3cm}
\section*{}
Neutrino oscillation search is one of the most promising way to go
beyond the Standard Model.  The existence of neutrino oscillation is
suggested by the lack of solar neutrino flux \cite{Ga1,Ga2,Kam,Cl} and
the anomalies of atmospheric neutrinos
\cite{AtmKam,IMB,SOUDAN2}~%
\footnote{%
Some experiments have not observed the atmospheric neutrino anomaly
\cite{NUSEX,Frejus}.%
}%
.  Further confirmation of its existence is now close at hand owing to
the long baseline neutrino oscillation experiments, such as K2K
experiments \cite{K2K} and Minos experiments \cite{Minos}.  The
possible effects of $CP$ and $T$ violation in such experiments have
been discussed in the literatures \cite{AS,Sato,AKS,MN,BGG,KS}.
There the matter density is approximated to be constant%
\footnote{%
We have checked numerically in Ref. \cite{KS} that the constant
density approximation works well for some parameter pace of neutrino's
in K2K experiments.  However it does not mean the approximation works
well for all the parameter space.  Another example of numerical
calculations is presented in Ref. \cite{Krastev}.%
}.  %
We now study effects of the matter density fluctuation and clarify
when the constant density approximation is valid.  We also verify that
this approximation works well for the K2K experiment.

Let $\nu = (\nu_{\rm e}, \nu_\mu, \nu_\tau)$ and $\nu' = (\nu_1,
\nu_2, \nu_3)$ be the eigenstates of flavor and mass in vacuum,
respectively.  They are related through the mixing matrix $U$ as
\begin{equation}
 \nu = U \nu'.
 \label{eq:mixing}
\end{equation}
Defining $\delta m^2_{ij} \equiv m^2_i - m^2_j$ with the mass
eigenvalues $m_i (i=1, 2, 3)$, the evolution equation of flavor
eigenstates in matter is given by
\begin{equation}
 {\rm i} \frac{{\rm d} \nu(x)}{{\rm d} x}
=
 H(x) \nu(x),
 \label{eq:evolution}
\end{equation}
\begin{equation}
 H(x)
\equiv
 \frac{1}{2E}
 \left\{
  U
  \left(
   \begin{array}{ccc}
    0& & \\
     &\delta m^2_{21}& \\
     & &\delta m^2_{31}
   \end{array}
  \right)
  U^{\dagger} +
  \left(
   \begin{array}{ccc}
    a(x)& & \\
     &0& \\
     & &0
   \end{array}
  \right)
 \right\}.
 \label{eq:Hdef}
\end{equation}
Here
\begin{equation}
 a(x)
 \equiv
 2 \sqrt{2} G_{\rm F} n_{\rm e}(x) E
 =
 7.56 \times 10^{-5} {\rm eV^2}
 \frac{\rho(x)}{\rm g~cm^{-3}} \frac{E}{\rm GeV},
 \label{eq:adef}
\end{equation}
$n_{\rm e}(x)$ is the electron density, $\rho(x)$ is the matter
density and $E$ is the neutrino energy.  The evolution equation
(\ref{eq:evolution}) is solved as
\begin{equation}
 \nu(L)
=
 {\rm T}\exp \left( - {\rm i} \int_0^L {\rm d} x H(x) \right)
 \nu(0)
\equiv
 S(L) \nu(0),
 \label{eq:ev-sol}
\end{equation}
giving the oscillation probability from $\nu_{\alpha}$ to $\nu_{\beta}
(\alpha, \beta = {\rm e}, \mu, \tau)$ as
\begin{equation}
 P(\nu_{\alpha} \rightarrow \nu_{\beta}; L, E)
 =
 \left| S(L)_{\beta \alpha} \right|^2.
 \label{eq:prob}
\end{equation}

We assume $\delta m^2_{31} \sim (10^{-2} \sim 10^{-3}) {\rm eV^2}$ and 
$\delta m^2_{21} \sim (10^{-5} \sim 10^{-4}) {\rm eV^2}$ allowing for
atmospheric neutrino anomaly and solar neutrino deficit.  Along with
$a \sim 10^{-4} {\rm eV^2}$ (see eq.(\ref{eq:adef})), we see
\begin{equation}
 \delta m^2_{21}, a \ll \delta m^2_{31}.
 \label{eq:llrel}
\end{equation}
We hence separate as $H(x)=H_0 + H_1(x)$, where
\begin{equation}
 H_0
\equiv
 \frac{1}{2E} U
 \left(
  \begin{array}{ccc}
  0& & \\
   &0& \\
   & &\delta m^2_{31}
  \end{array}
 \right)
 U^{\dagger}
 \label{eq:H0def}
\end{equation}
and
\begin{equation}
 H_1(x)
\equiv
 \frac{1}{2E}
 \left\{
  U
  \left(
   \begin{array}{ccc}
    0& & \\
     &\delta m^2_{21}& \\
     & &0
   \end{array}
  \right)
  U^{\dagger} +
  \left(
   \begin{array}{ccc}
    a(x)& & \\
     &0& \\
     & &0
   \end{array}
  \right)
 \right\},
 \label{eq:H1def}
\end{equation}
and treat $H_1(x)$ as a perturbation.  Taking up to the lowest order in
$H_1(x)$, we obtain
\begin{eqnarray}
 S(L)
&\simeq&
 {\rm e}^{ -{\rm i} H_0 L }
+
 {\rm e}^{ -{\rm i} H_0 L } (-{\rm i})
 \int_0^L {\rm d} x
 \left[
  {\rm e}^{{\rm i} H_0 x} H_1(x) {\rm e}^{-{\rm i} H_0 x}
 \right]
 \nonumber \\
&\equiv&
 S_0(L) + S_1(L).
 \label{eq:S0&S1}
\end{eqnarray}
The fluctuation of matter density $\rho(x)$, or equivalently that of
$a(x)$, affects $S_{1}(L)$ alone.  We separate that effect from
$S_{1}(L)$ in the following.

Letting
\begin{equation}
 \bar a \equiv \frac{1}{L} \int_0^L {\rm d} x \; a(x)
 \label{eq:a-bar-def}
\end{equation}
and
\begin{equation}
 \delta a(x) \equiv a(x) - \bar a,
 \label{eq:delta-a-def}
\end{equation}
we separate the contribution of $\delta a(x)$ from $H_1(x)$ by
defining
%
\begin{equation}
 \bar H_1(x)
\equiv
 \frac{1}{2E}
 \left\{
  U
  \left(
   \begin{array}{ccc}
    0& & \\
     &\delta m^2_{21}& \\
     & &0
   \end{array}
  \right)
  U^{\dagger} +
  \left(
   \begin{array}{ccc}
    \bar a& & \\
     &0& \\
     & &0
   \end{array}
  \right)
 \right\}
 \label{eq:H1-bar-def}
\end{equation}
and
\begin{equation}
 \delta H_1(x)
 \equiv
 \frac{1}{2E}
  \left(
   \begin{array}{ccc}
    \delta a(x)& & \\
     &0& \\
     & &0
   \end{array}
  \right).
 \label{eq:delta-H1-def}
\end{equation}
Defining accordingly
\begin{equation}
 \bar S_1(L)
=
 -{\rm i} \int_0^L {\rm d} x
 \left[
  {\rm e}^{ -{\rm i} H_0 (L-x)} \bar H_1(x) {\rm e}^{-{\rm i} H_0 x}
 \right]
 \label{eq:S1bar-def}
\end{equation}
and
\begin{eqnarray}
 \delta S_1(L)
=
 -{\rm i} \int_0^L {\rm d} x
 \left[
  {\rm e}^{ -{\rm i} H_0 (L-x)} \delta H_1(x) {\rm e}^{-{\rm i} H_0 x}
 \right],
 \label{eq:delta-S1-def}
\end{eqnarray}
$\bar S_1$ is calculated to be \cite{AKS}
\begin{eqnarray}
 \bar S_1(x)_{\beta \alpha}
&=&
 - {\rm i} U_{\beta 2} U^{\ast}_{\alpha 2}
 \cdot \frac{\delta m^2_{21} x}{2E}
 \nonumber \\
& &
 - {\rm i}
 (\delta_{\beta {\rm e}} - U_{\beta 3} U^{\ast}_{ {\rm e} 3 })
 (\delta_{\alpha {\rm e}} - U_{ {\rm e} 3 } U^{\ast}_{\alpha 3})
 \cdot \frac{\bar a x}{2E}
 \nonumber \\
& &
 + U_{\beta 3} U^{\ast}_{ {\rm e} 3 }
 (\delta_{\alpha {\rm e}} - U_{ {\rm e} 3 } U^{\ast}_{\alpha 3})
 \left[ \exp \left( - {\rm i} \frac{\delta m^2_{31} x}{2E} \right) - 1 \right]
 \cdot \frac{\bar a}{\delta m^2_{31}}
 \nonumber \\
& &
 + (\delta_{\beta {\rm e}} - U_{\beta 3} U^{\ast}_{ {\rm e} 3 })
 U_{ {\rm e} 3 } U^{\ast}_{\alpha 3}
 \left[ \exp \left( - {\rm i} \frac{\delta m^2_{31} x}{2E} \right) - 1 \right]
 \cdot \frac{\bar a}{\delta m^2_{31}}
 \nonumber \\
& &
 - {\rm i} 
 U_{\beta 3}U^{\ast}_{\alpha 3} \left| U_{ {\rm e} 3 } \right|^2
 \exp \left( - {\rm i} \frac{\delta m^2_{31} x}{2E} \right)
 \cdot \frac{\bar a x}{2E}.
 \label{eq:S1long}
\end{eqnarray}
To evaluate $\delta S_1 (x)_{\beta \alpha}$ we expand $\delta a(x)$
as
\begin{equation}
 \delta a(x)
\equiv
 \sum_{n=-\infty}^{\infty}
  a_n {\rm e}^{ -{\rm i} n \pi x / L }
\end{equation}
and carry out integration for each Fourier component.  Note that $a_0
= 0$ from the definition (\ref{eq:delta-a-def}).  We then find
\begin{eqnarray}
 \delta S_1(x)_{\beta \alpha}
&=&
 \delta_{\alpha {\rm e}} U_{\beta 3} U^\ast_{ {\rm e} 3 }
 \left[ \exp \left( - {\rm i} \frac{\delta m^2_{31} x}{2E} \right) - 1 \right]
 \cdot
 \sum_{n\neq0}
 \frac{a_n}{\delta m^2_{31}}
  \left(
   1 - \frac{2 n \pi}{\delta m^2_{31} x / 2E}
  \right)^{-1}
 \nonumber \\
&+&
 \delta_{\beta {\rm e}} U_{ {\rm e} 3} U^\ast_{\alpha 3}
 \left[ \exp \left( - {\rm i} \frac{\delta m^2_{31} x}{2E} \right) - 1 \right]
 \cdot
 \sum_{n\neq0}
 \frac{a_n}{\delta m^2_{31}}
  \left(
   1 + \frac{2 n \pi}{\delta m^2_{31} x / 2E}
  \right)^{-1}
 \nonumber \\
&-&
 U_{\beta 3} U^\ast_{\alpha 3} \left| U_{ {\rm e} 3} \right|^2
 \left[ \exp \left( - {\rm i} \frac{\delta m^2_{31} x}{2E} \right) - 1 \right]
 \times
 \nonumber \\
 & &
 \sum_{n\neq0}
 \frac{a_n}{\delta m^2_{31}}
 \left[
  \left(
   1 - \frac{2 n \pi}{\delta m^2_{31} x / 2E}
  \right)^{-1}
  +
  \left(
   1 + \frac{2 n \pi}{\delta m^2_{31} x / 2E}
  \right)^{-1}
 \right].
 \label{eq:deltaS1:2}
\end{eqnarray}

We now see the order of magnitude of $\bar S_1$ and $\delta S_1(x)$
((\ref{eq:S1long}) and (\ref{eq:deltaS1:2})).  Here we assume that the
products of $U$'s and $\delta m^2_{31} L / 2E$ are all $O(1)$ so that
we can observe the neutrino oscillation.  In this case we can see from
(\ref{eq:S1long}) that
\begin{equation}
 \bar S_1 =
 O \left(
  \frac{\delta m^2_{21}}{\delta m^2_{31}}
  \quad \mbox{or} \quad
  \frac{\bar a}{\delta m^2_{31}}
 \right).
\end{equation}
On the other hand all the three terms of $\delta S_1(x)$ contains
factors $( a_n / \delta m^2_{31} )$ and also $\{ 1 - 2 n \pi / (\delta
m^2_{31} / 2E) \}^{-1} \sim 1/n$.  Hence
\begin{equation}
 \delta S_1(x) =
 O \left( \frac1n \frac{a_n}{\bar a} \bar S_1 \right).
 \label{eq:deltaS1order}
\end{equation}

We can see from (\ref{eq:deltaS1order}) that the long wavelength
fluctuation (i.e. small $n$) of the matter density is important; short
wavelength fluctuations (i.e. large $n$) of matter density is in
general irrelevant due to the factor $1/n$.  This means we do not need
to survey detailed profile of the matter density distribution on the
baseline.  On the other hand we must check on a case-by-case basis whether
the long wavelength fluctuation is relevant or not.  This check can be 
done by considering the magnitudes of $a_n / \bar a$.

\begin{figure}
 \unitlength=1cm
  \begin{picture}(15,7)
  \unitlength=1mm
  \centerline{
   \epsfxsize=13cm
   \epsfbox{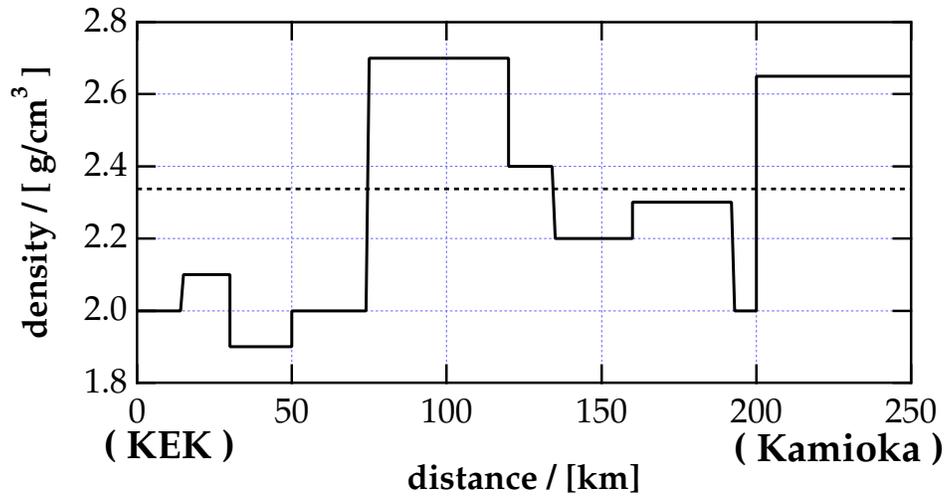}
   } 
\end{picture}
\caption[K2Kdensity]
{%
The density profile between KEK and Kamioka.
}%
 \label{K2Kdensity}
\end{figure}

\begin{figure}
 \unitlength=1cm
  \begin{picture}(15,7)
  \unitlength=1mm
  \centerline{
   \epsfxsize=13cm
   \epsfbox{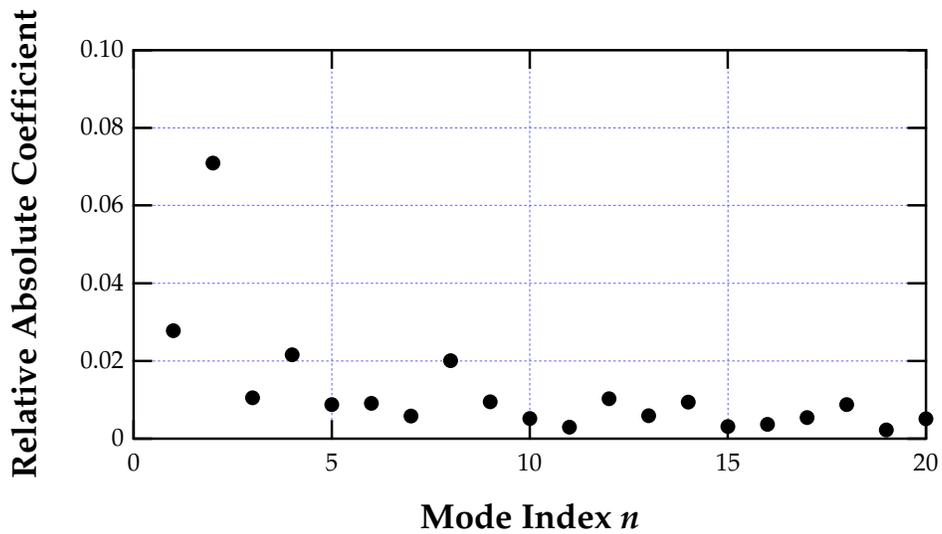}
   } 
\end{picture}
\caption[relcoeff]
{%
The values of $a_n / \bar a$ for first several $n$'s.
}%
 \label{relcoeff}
\end{figure}

Let us carry out a check for the K2K experiments as an example.
Figure \ref{K2Kdensity} shows the density profile between KEK and
Kamioka \cite{Koma}.  The first several Fourier coefficients divided
by the mean density ($= a_n / a$) for this profile is given in
Fig. \ref{relcoeff}.
We can see that $a_n / \bar a$ is small, mostly much less than 0.1.
This justifies the constant matter density approximation for K2K
experiments.

We have shown that short wavelength fluctuation of matter density is
irrelevant and only first several Fourier coefficients of fluctuation
may be important for long baseline neutrino oscillation experiments.
Since we can expect that such fluctuations are much smaller than the
mean density in crust we can approximate the matter density to be
constant.  This is indeed the case for K2K experiments.  We could
check the approximation works
well independent of the parameters relevant to neutrino oscillation.%
\section*{Acknowledgment}
The authors thank Professor J.~Arafune for valuable discussions and
encouragement.  One of the authors (J.S.) is JSPS Fellowship.
%


\begin{thebibliography}{99}

\bibitem{Ga1} GALLEX Collaboration, P.~Anselmann {\it et al.}, Phys.
    Lett. B {\bf 357} (1995) 237.
    
\bibitem{Ga2} SAGE Collaboration, J.~N.~Abdurashitov {\it et al.},
    Phys. Lett. B {\bf 328} (1994) 234.
    
\bibitem{Kam} Kamiokande Collaboration, Y.~Suzuki, Nucl. Phys. B (Proc.
    Suppl.) {\bf 38} (1995) 54.
    
\bibitem{Cl} Homestake Collaboration, B.~T.~Cleveland {\it et al.},
    Nucl. Phys. B (Proc. Suppl.) {\bf 38} (1995) 47.

\bibitem{AtmKam} Kamiokande Collaboration, K.~S.~Hirata {\it et al.},
  Phys. Lett. B {\bf 205} (1988) 416; {\it ibid.} B {\bf 280} (1992)
  146; Y.~Fukuda {\it et al.}, Phys. Lett. B {\bf 335}, 237 (1994).
    
\bibitem{IMB} IMB Collaboration, D.~Casper {\it et al.},
  Phys. Rev. Lett. {\bf 66} (1991) 2561;\\ R.~Becker-Szendy {\it et
  al.}, Phys. Rev. D {\bf 46} (1992) 3720.

\bibitem{SOUDAN2} SOUDAN2 Collaboration, T.~Kafka, Nucl. Phys. B
  (Proc.  Suppl.) {\bf 35} (1994) 427; M.~C.~Goodman, {\it ibid.} {\bf
  38} (1995) 337; W.~W.~M.~Allison {\it et al.}, Phys. Lett. B {\bf
  391} (1997) 491.
    
\bibitem{NUSEX} NUSEX Collaboration, M.~Aglietta {\it et al.},
  Europhys. Lett. {\bf 8} (1989) 611; {\it ibid.} {\bf 15} (1991) 559.
    
\bibitem{Frejus} Fr\'ejus Collaboration, K.~Daum {\it et al.},
  Z. Phys. C {\bf 66} (1995) 417.

\bibitem{K2K} K.~Nishikawa, INS-Rep-924 (1992).
    
\bibitem{Minos} S.~Parke, Fermilab-Conf-93/056-T (1993)
  (hep-ph/9304271).

\bibitem{AS} J.~Arafune and J.~Sato, Phys. Rev. D {\bf 55} (1997)
  1653.

\bibitem{Sato} J.~Sato, Nucl. Phys. B (Proc. Suppl.) {\bf 59} (1997)
  262.
  
\bibitem{AKS} J.~Arafune, M.~Koike and J.~Sato, Phys. Rev. D {\bf 56}
  (1997) 3093.

\bibitem{MN} H.~Minakata and H.~Nunokawa, Phys. Lett. B {\bf 413}
  (1997) 369.

\bibitem{BGG} S.~M.~Bilenky, G.~Giunti and W.~Grimus, UWThPh-1997-51,
  DFTT 74/97 (hep-ph/9712537).

\bibitem{KS} M.~Koike and J.~Sato, in {\it Proceedings of the 5th KEK
  Meeting on CP Violation and its Origin} (ed. K. Hagiwara), KEK
  Proceedings 97-12.

\bibitem{Krastev} P.~I.~Krastev, Nuovo Cim. 103A (1990) 361.

\bibitem{Koma} M.~Komazawa, Private Communication.
\end{thebibliography}
\end{document}